\documentclass[superscriptaddress,preprint,amsmath,amssymb,aps,pre]{revtex4-2}

\usepackage{graphicx}
\usepackage{amsmath}
\usepackage{dcolumn}
\usepackage{txfonts}
\usepackage{bm}
\usepackage{mathrsfs}
\usepackage{amsfonts}
\usepackage{amssymb}
\usepackage{appendix}
\usepackage{braket}
\usepackage{color}
\usepackage[colorlinks=true,citecolor=magenta,anchorcolor=blue]{hyperref}

\begin{document}

\title{Convection-modulated topological edge mode and extended-localized criticality in thermal metamaterials}

\author{Zhoufei Liu}
\author{Jiping Huang}\email{jphuang@fudan.edu.cn}
\affiliation{Department of Physics, State Key Laboratory of Surface Physics, and Key Laboratory of Micro and Nano Photonic Structures (MOE), Fudan University, Shanghai 200438, China}

\author{Ying Li}\email{eleying@zju.edu.cn}
\affiliation{State Key Laboratory of Extreme Photonics and Instrumentation, ZJU-Hangzhou Global Scientific and Technological Innovation Center, Zhejiang University, Hangzhou 310027, China}
\affiliation{International Joint Innovation Center, Zhejiang Key Laboratory of Intelligent Electromagnetic Control and Advanced Electronic Integration, The Electromagnetics Academy of Zhejiang University, Zhejiang University, Haining 314400, China}
\affiliation{Shaoxing Institute of Zhejiang University, Zhejiang University, Shaoxing 312000, China}

\date{\today}

\begin{abstract}

Convection offers a dynamic and flexible approach to achieving a variety of novel physical phenomena beyond pure conduction. Here, we demonstrate that thermal metamaterials with convection modulation enable the realization of non-Hermitian topological edge modes and bulk mode criticality. We illustrate that a periodic modulation can induce localized edge modes within the band gap. The temperature field of the topological state is localized at the edge rings, decaying exponentially at a fixed rate. Additionally, we introduce an extended-localized criticality through the quasiperiodic convection modulation in thermotics. The convections have an advantage of fantastic   tunability in the application. Our work proposes a scheme for implementing topological modes and bulk mode criticality through modulating convection in diffusion systems, paving the way for the design of reconfigurable thermal devices.

\end{abstract}

\maketitle

\section{\label{sec:introduction}Introduction}

Over the past few decades, topological insulators have emerged as pivotal platforms for exploring novel phenomena in condensed matter physics~\cite{HasanRMP10, QiRMP11}. A distinctive feature of these insulators is the presence of robust edge states, which can be characterized by the bulk band. Beyond real materials, topological phases can also be realized in various artificial metamaterials, including photonics~\cite{LuNP14, OzawaRMP19}, acoustics~\cite{XueNRM22}, and mechanics~\cite{HuberNP16}. Except for the periodic lattice, topological states can also arise in the quasiperiodic lattice, exemplified by the Aubry-André-Harper (AAH) model~\cite{Aubry80, Harper55}. The one-dimensional AAH model can be mapped onto a two-dimensional quantum Hall insulator~\cite{LangPRL12, KrausPRL12, VerbinPRL13}. Furthermore, the quasiperiodic model can exhibit the extended-localized criticality at a finite transition point~\cite{RoatiNat08, LahiniPRL09}.

Recently, the flourishing of non-Hermitian physics has invigorated the study of topological states, leading to numerous exotic phenomena~\cite{AshidaAP20, BergholtzRMP21}. In addition to the non-Hermitian skin effect~\cite{YaoPRL18, ZhangAP22, LinFP23}, topological edge modes induced by gain and loss represent an intriguing topic without a Hermitian counterpart~\cite{TakataPRL18, LuoPRL19, YuNat24}. On photonic platforms, gain and loss can be readily implemented and finely tuned in experiments. Consequently, topological states achieved through this method facilitate the design of reconfigurable photonic devices~\cite{ZhaoSci19, BogaertsNat20, LopezNC20, OnNC24}. Furthermore, non-Hermitian quasiperiodic lattices (such as the non-Hermitian AAH model) exhibit a range of novel topological properties and localization behaviours~\cite{LonghiPRL19, LonghiPRB19, JiangPRB19, LiuPRB21-1, LiuPRB21-2, WangPRL20, ZhouPRL23, ZhuPRR23}. Notably, the quasiperiodic gain and loss modulation can engineer the extended-localized criticality of bulk modes~\cite{PereiraPRR24}.

Thermal metamaterials are synthetic structural materials capable of independently regulating heat flow~\cite{YangPR21, LiNRM21, ZhangNRP23, YangRMP24}. In recent years, they have emerged as a novel platform for realizing topological phases, thereby fostering a burgeoning interdisciplinary research field termed ``topological thermotics''~\cite{LiuNRP24, XuIJHMT21, YoshidaSR21}. However, the majority of reported thermal topological states are rooted in static structures~\cite{QiAM22, HuAM22, LiuPRL24, WuAM23}, limiting their adjustability. As a dynamically tunable parameter, convection has been incorporated into thermal metamaterial research to elicit a variety of novel functionalities and phenomena~\cite{JuAM22}, including enhanced effective thermal conductivity~\cite{LiNM19, XuNC20}, spatiotemporal nonreciprocity~\cite{XuPRL22-1, XuPRL22-2, LiNC22}, and notably, non-Hermitian physics~\cite{LiSci19, XuPRL21, CaoSA24, XuPNAS22}. Leveraging its wave-like behaviour, convection can mimic the role of gain and loss in the wave propagation within naturally dissipative diffusion processes. Despite extensive researches in wave systems, the manipulation of convection to achieve topological modes and bulk mode criticality in diffusion systems remains unexplored.

In this study, we showcase the realization of topological edge modes and bulk mode criticality through convection in thermal diffusion. Our theoretical model, incorporating a modulated imaginary onsite potential, is implemented using a coupled ring chain structure. By periodically and quasiperiodically modulating the rotating velocities of the rings, we demonstrate the topological edge mode and extended-localized transition in thermal metamaterials, respectively. We perform the temperature field simulation to analyze their thermal behaviours. A recent work has excellently demonstrated the topological edge mode and extended-localized criticality with engineered gain and loss in photonics~\cite{PereiraPRR24}. However, due to the dissipative nature of thermal diffusion, the temperature field will exhibit a completely different behaviour with the wave propagation. The temperature field for edge mode decays fast and exponentially with a localized profile. In the extended phase, the temperature field exhibits a uniform and stationary profile. While in the localized phase, the thermal field features a mobile multiple localization centers phenomenon. Our findings provide a blueprint for designing reconfigurable thermal devices based on tunable convection. 

The paper is structured as follows: In Sec.~\ref{sec:coupled_ring}, we introduce the theoretical model with engineered loss modulation and demonstrate its implementation in thermotics using the coupled ring chain structure. In Sec.~\ref{sec:topological}, we delve into the topological edge mode driven by periodic convection modulation. In Sec.~\ref{sec:extend_local}, we investigate the extended-localized criticality of bulk modes with a quasiperiodic modulation. Finally, we conclude our findings in Sec.~\ref{sec:conclusion}.

\section{\label{sec:coupled_ring}Theoretical model and coupled ring chain structure}

\begin{figure}
\includegraphics[width=0.7\linewidth]{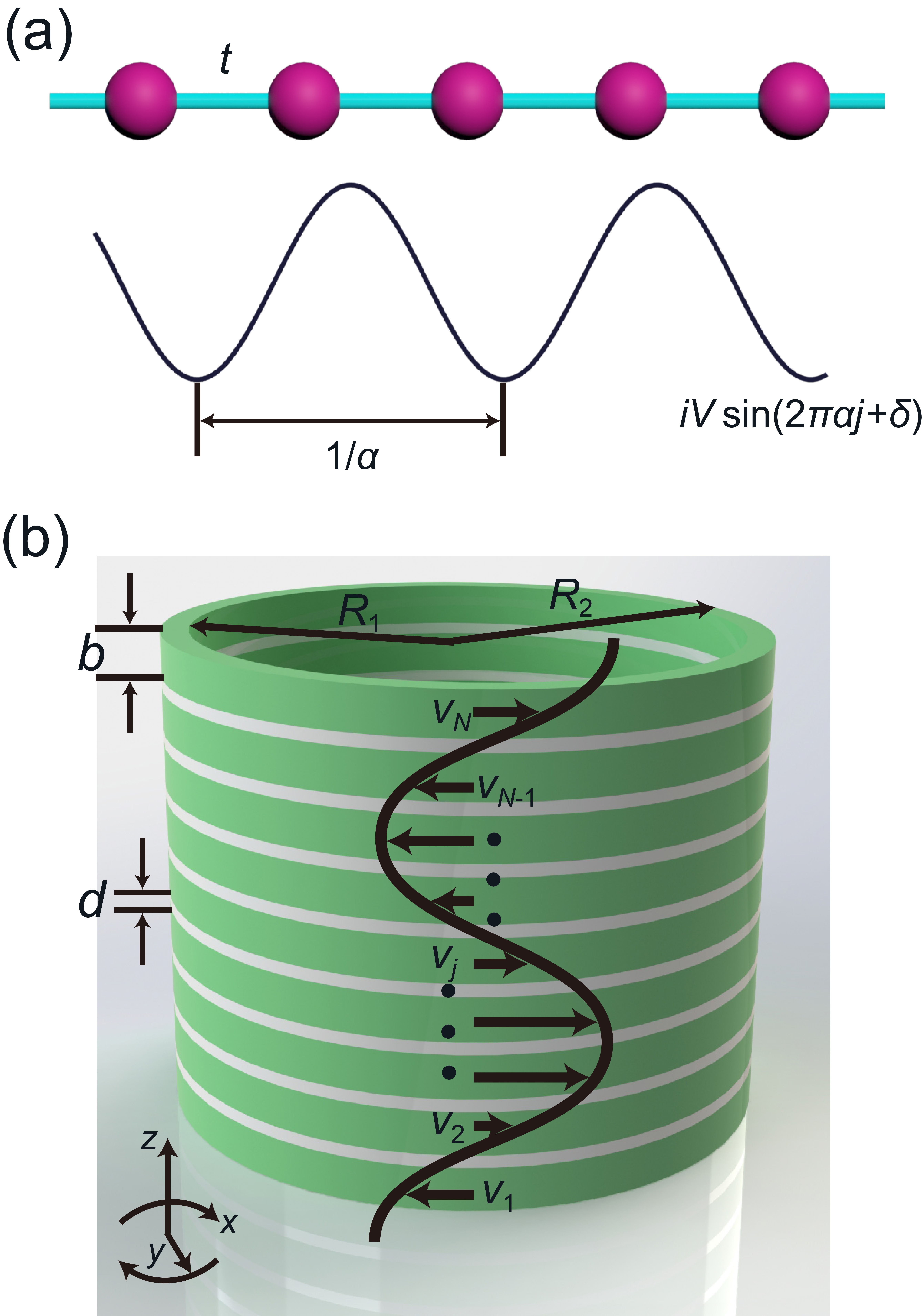}
\caption{Tight-binding model and coupled ring chain structure. (a) Schematic diagram of the one-dimensional lattice with imaginary onsite modulation. The theoretical model has the uniform nearest-neighbour hopping $t$ (upper panel), and an imaginary onsite potential $iV_{j}$ (lower panel), which varies sinusoidally in space with the period $1/{\alpha}$. (b) Schematic diagram of the coupled ring chain structure. The green area denotes the ring, whose thickness is denoted as $b$. The grey area represents the interlayer, whose thickness is denoted as $d$. The inner and outer radii of rings and interlayers are denoted as $R_{1}$ and $R_{2}$ with an approximation $R{\approx}R_{1}{\approx}R_{2}$. Here, the bottom channel is labelled as the first ring, while the top channel is denoted as the $N$-th ring. For the $j$-th ring, the rotation velocity is set as $v_{j}=RV{\,}{\rm{sin}}(2{\pi}{\alpha}j+{\delta})$.}
\label{Fig1}
\end{figure}

We investigate a one-dimensional model characterized by a uniform hopping amplitude $t$ between neighbouring sites and a purely imaginary onsite potential, which follows a sinusoidal modulation, as illustrated in Fig.~\ref{Fig1}(a). The Hamiltonian for this model is expressed as:
\begin{equation}\label{1}
\hat{H}=t\sum\limits_{j=1}^{N-1}(a_{j}^{\dagger}a_{j+1}+{\rm{H.c.}})+i\sum\limits_{j=1}^{N}V_{j}a_{j}^{\dagger}a_{j},
\end{equation}
where $N$ denotes the total number of sites, $a_{j}^{\dagger}$ ($a_{j}$) is the creation (annihilation) operator at the $j$-th site, and the corresponding onsite potential $V_{j}$ is 
\begin{equation}\label{2}
V_{j}=V{\,}{\rm{sin}}(2{\pi}{\alpha}j+{\delta}),
\end{equation}
where $V$, $\alpha$, and $\delta$ are the amplitude, inverse period, and phase of the onsite potential, respectively. This model serves as a non-Hermitian extension of the Hermitian AAH model~\cite{Aubry80, Harper55}. When $\alpha$ is a rational number $p/q$ (where $p$ and $q$ are coprime integers), the commensurate lattice exhibits a periodicity of $q$ sites. Specifically, if $q$ is a multiple of four, topological edge modes with purely imaginary eigenvalues emerge, protected by the non-Hermitian particle-hole symmetry~\cite{ZhuPRR23}. A special case is $\alpha=1/4$, featuring a repetitive gain/loss sequence, which has been explored by Takata and Notomi~\cite{TakataPRL18}. In this scenario, the lattice hosts only topological edge modes at zero real eigenvalues. Conversely, for other rational values of $\alpha$, topological edge modes appear at finite real energies within different band gaps. When $\alpha$ is irrational (such as the inverse golden ratio), the system adopts a quasiperiodic nature, yielding a fractal energy spectrum with topological edge modes residing in the band gap. Notably, this quasiperiodic modulation demonstrates an extended-localized transition of bulk modes~\cite{PereiraPRR24}. Next, we want to realize these intriguing phenomena in thermal metamaterials.

The basic thermal setup employed in this study is the coupled ring chain structure~\cite{CaoCP21, CaoCPL22, LiuPRAP24, LiuCPL23}, as illustrated in Fig.~\ref{Fig1}(b). This structure comprises several identical rings vertically interconnected through interlayer along the $z$-axis to form a chain. Applying Fourier's law of heat transfer, the thermal coupling equation for the $j$-th ring can be formulated as:  
\begin{equation}\label{3}
\frac{\partial T_{j}(x,t)}{\partial t}=\frac{\kappa}{\rho C}\frac{\partial^2 T_{j}(x,t)}{\partial x^2}+v_j\frac{\partial T_{j}(x,t)}{\partial x}+h\left[T_{j-1}(x,t)-T_{j}(x,t)\right]+h\left[T_{j+1}(x,t)-T_{j}(x,t)\right],
\end{equation} 
Here, $\kappa$, $\rho$, and $C$ represent the thermal conductivity, mass density, and heat capacity of the rings, respectively. The position along the ring is denoted by $x$. The heat exchange coefficient between the interlayer and the ring is given by $h=\kappa_{\mathrm{I}}/(\rho C b d)$, where $\kappa_{\mathrm{I}}$ is the interlayer's thermal conductivity. Furthermore, $T_{j}(x,t)$ and $v_j$ signify the temperature field and the modulated rotating velocity of the $j$-th ring, respectively. Given the periodic nature of the structure, we postulate that Eq.~\ref{3} has a plane wave solution of the form $T_{j}(x,t)=A_{j}e^{i(\beta x-\omega t)}$. Here, $A_j$ denotes the amplitude of the temperature field for the $j$-th ring, and $\omega$ represents the eigenfrequency (the imaginary part of $\omega$ is decay rate). The propagation constant $\beta$ is defined as $\beta=2m\pi/L=m/R$, where $m$ is the mode order. This study focuses on the fundamental mode ($m=1$), as only the slowest decaying mode can be observed in diffusion systems. Substituting the plane wave solution into Eq.~\ref{3}, we derive the effective Hamiltonian of the coupled ring chain structure under the open boundary condition, expressed in the second quantized form as:
\begin{equation}\label{4}
\hat{H}=ih\sum\limits_{j=1}^{N-1}(a_{j}^{\dagger}a_{j+1}+{\rm{H.c.}})+i\sum\limits_{j=1}^{N}(S+i{\beta}v_{j})\hat{a}_{j}^{\dagger}\hat{a}_{j},
\end{equation}
where $i=\sqrt{-1}$ is the imaginary unit, and $S=-\left[{\beta}^{2}{\kappa}/({\rho C})+2h\right]$ is a constant onsite term. By modulating each ring's rotating velocity (i.e., thermal convection) as $v_{j}=RV{\,}{\rm{sin}}(2{\pi}{\alpha}j+{\delta})$, the coupled ring chain Hamiltonian Eq.~\ref{4} has the same form with the model Hamiltonian Eq.~\ref{1}, differing only by a factor of $i$. Consequently, the coupled ring chain structure exhibits the thermal behaviour of the theoretical model.

\section{\label{sec:topological}Convection-modulated topological edge mode}

\begin{figure}
\includegraphics[width=0.9\linewidth]{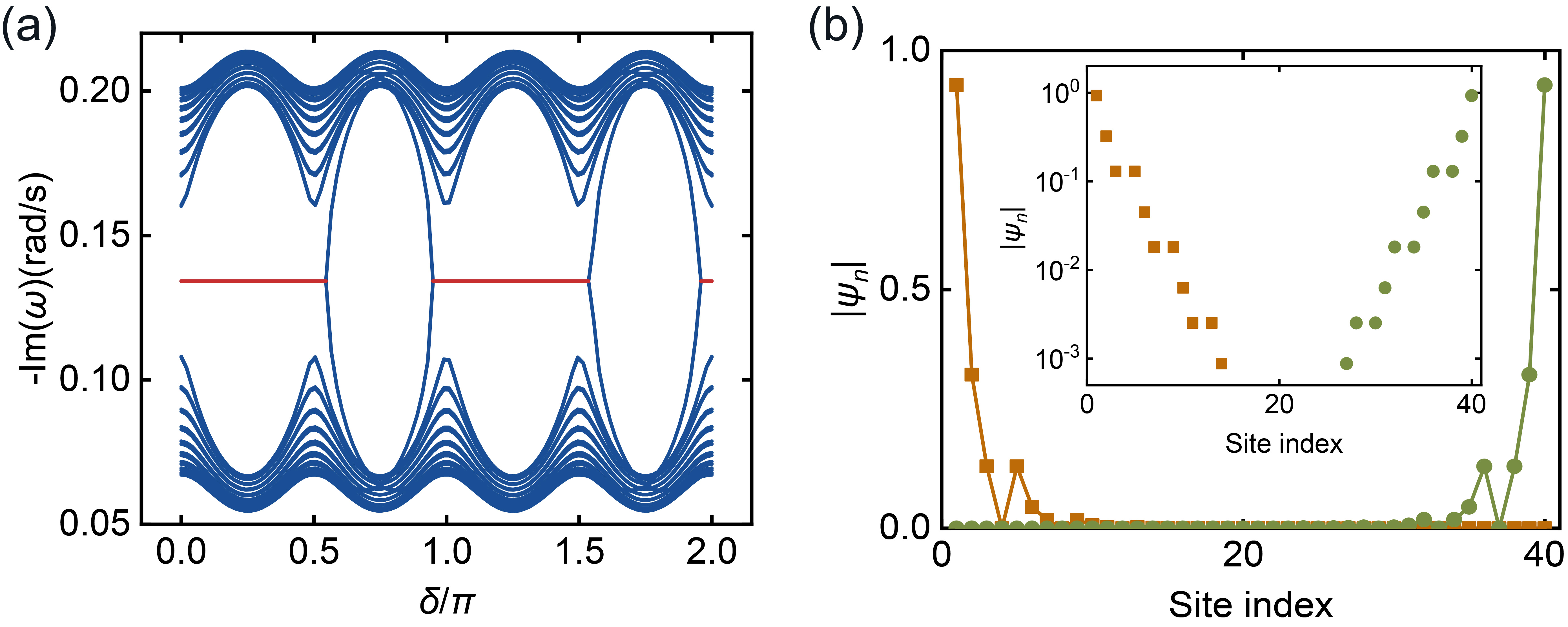}
\caption{Eigenvalue and eigenstate of diffusive model when $\alpha=1/4$. (a) The theoretical imaginary spectrum with different addition phases $\delta$ under the open boundary condition. The red lines mark the topological edge modes. (b) The theoretical eigenstate distributions of two topological edge modes at $\delta=0.25\pi$. The inset is the semilogarithmic plot for the distribution of two edge modes. Here $V=2h$.}
\label{Fig2}
\end{figure}

In this section, we explore the topological edge modes arising from convection modulation. Here we set $\alpha$ to $1/4$. Initially, we analyze the results of the diffusive theoretical model. The imaginary part of the spectrum under the open boundary condition reveals topological edge states within the band gap, depicted as red curves in Fig.~\ref{Fig2}(a). As the phase $\delta$ varies, a topological phase transition occurs at the gap closing point. The wave functions of these two topological edge modes are exponentially localized at the boundary, as illustrated in Fig.~\ref{Fig2}(b). For this model, the band topology is characterized by the non-Hermitian generalization of electric polarization, calculated using the non-Abelian Wilson loop method (see Appendix~\ref{app:AppA} and Fig.~\ref{Fig7}). In the nontrivial phase, the polarization is quantized as $|p_{x}|=1/2$, whereas it vanishes in the trivial phase.

\begin{figure}
\includegraphics[width=0.9\linewidth]{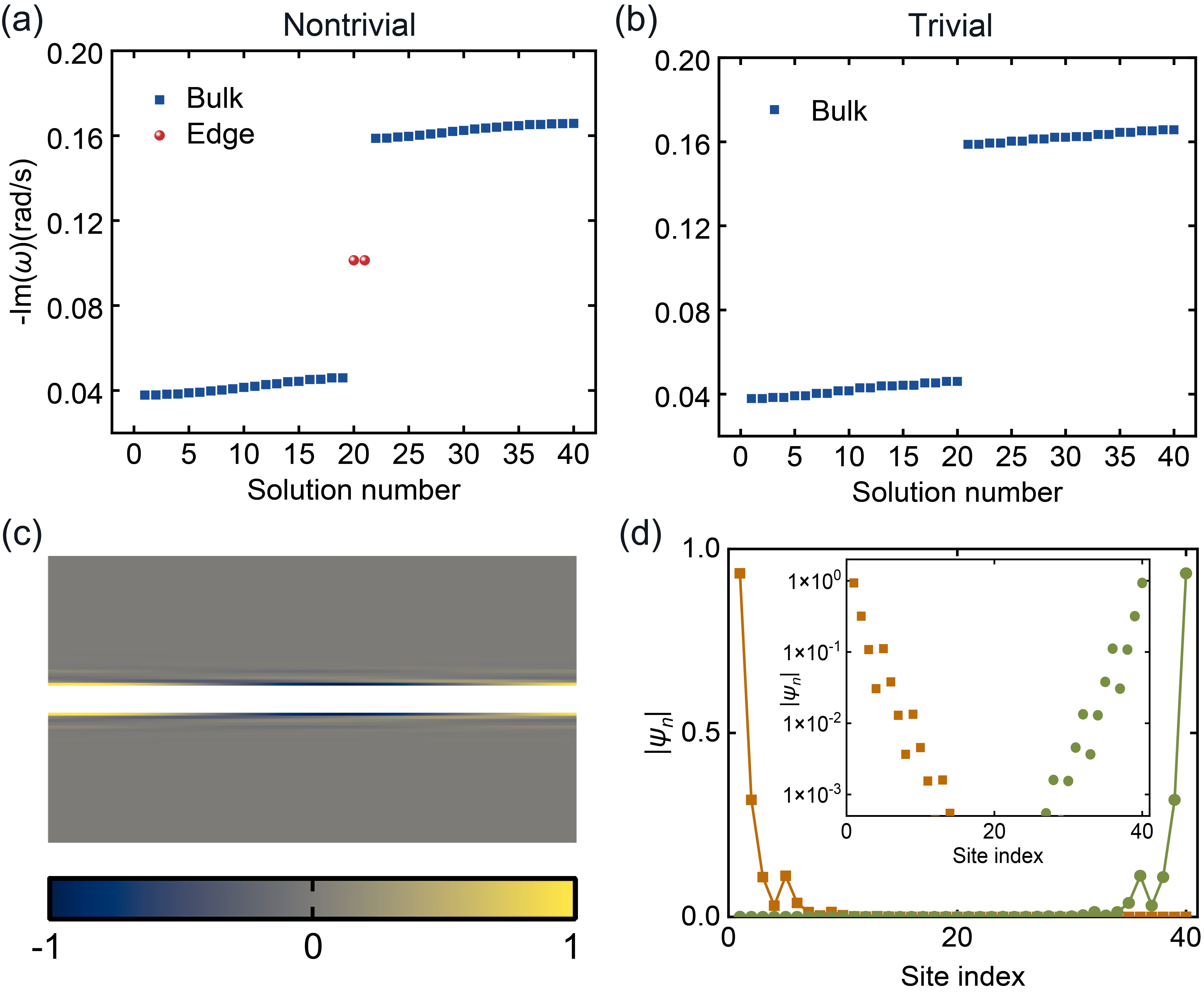}
\caption{Eigenvalue and eigenstate of simulated structure when $\alpha=1/4$. (a,b) The simulated imaginary spectra at (a) $\delta=0.25\pi$ and (b) $\delta=0.75\pi$. The blue square and red sphere indicate the bulk and edge modes, respectively. (c) The simulated eigenfield profiles of two topological edge modes at $\delta=0.25\pi$. (d) The corresponding simulated eigenstate distributions of two topological edge modes at $\delta=0.25\pi$. The inset is the semilogarithmic plot for the distribution of two topological edge modes.}
\label{Fig3}
\end{figure}

Then we conduct the band structure simulation for the coupled ring chain structure (see the method in Appendix~\ref{app:AppB}). In the topological nontrivial phase ($\delta=0.25\pi$), two topological edge modes emerge within the imaginary band gap, as shown in Fig.~\ref{Fig3}(a). Furthermore, these edge modes are robust against disorder of the rotating velocities of rings [see Appendix~\ref{app:AppC} and Fig.~\ref{Fig8}(a)]. Conversely, in the trivial phase ($\delta=0.75\pi$), no edge state is observed in the imaginary spectrum, as depicted in Fig.~\ref{Fig3}(b). From the eigenstate simulation results, we observe that the two topological edge modes are localized at opposite boundaries [see Fig.~\ref{Fig3}(c)] and exhibit exponential localization behaviour [see Fig.~\ref{Fig3}(d)]. Furthermore, in the case without convection modulation, there is no band gap in the imaginary spectrum and the eigenstate distributions are extended [see Appendix~\ref{app:AppE} and Figs.~\ref{Fig10}(a,b)]. Therefore, we conclude that the coupled ring chain structure with convection modulation accurately captures the properties of the diffusive theoretical model.

\begin{figure}
\includegraphics[width=0.9\linewidth]{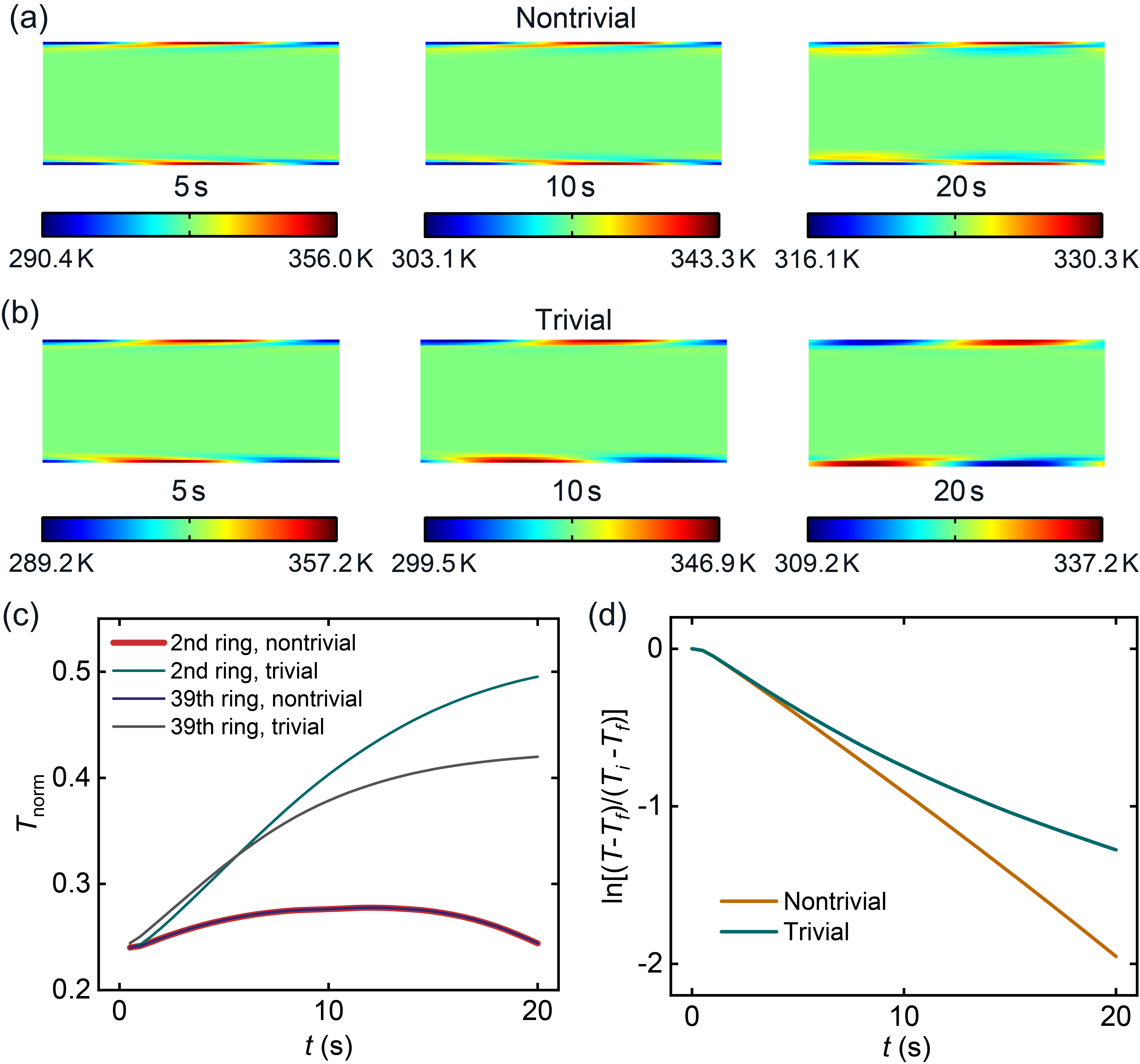}
\caption{Temperature field simulation with $\alpha=1/4$. (a) represents the nontrivial phase ($\delta=0.25\pi$), while (b) depicts the trivial phase ($\delta=0.75\pi$). (c) The normalized temperature field $T_{\rm{norm}}$ for the 2nd and 39th rings with time evolution in the nontrivial and trivial phases. Here the normalized temperature is $T_{\rm{norm}}=(T_{\rm{target,max}}-T_{\rm{target,min}})/\sqrt{\sum\limits_{j=1}^{N}(T_{j,{\rm{max}}}-T_{j,{\rm{min}}})^2}$, where $T_{\rm{target,max}}$ and $T_{\rm{target,min}}$ ($T_{j,{\rm{max}}}$ and $T_{j,{\rm{min}}}$) are the maximum and minimum temperatures for the targeted ($j$-th) ring.  (d) The normalized maximum temperature evolution ${\rm{ln}}[(T-T_{f})/(T_{i}-T_{f})]$ of the edge (1st or 40th) ring in the nontrivial and trivial phases.}
\label{Fig4}
\end{figure}

Next, we perform the temperature field simulation to investigate the thermal behaviour of topological edge modes (see the method in Appendix~\ref{app:AppB}). We select the mode profiles of two edge modes at $\delta=0.25\pi$ as the excited temperature patterns. During time evolution, the temperature field in the nontrivial phase [see Fig.~\ref{Fig4}(a)] exhibits a stronger localization at the edge rings compared to the trivial phase [see Fig.~\ref{Fig4}(b)]. Specifically, in the nontrivial phase, the edge ring introduces less heat to its adjacent ring. For a quantitative analysis, we extract the normalized temperature field for the 2nd and 39th rings, as shown in Fig.~\ref{Fig4}(c). In the nontrivial phase with edge modes, the normalized temperature field remains small and changes slightly over time. Conversely, in the trivial case without edge mode, the normalized temperature field is relatively larger and increases with time obviously, indicating a greater amount of heat transfer from the 1st (40th) ring to the 2nd (39th) ring. We then examine the maximum temperature evolution of the edge ring. In the nontrivial phase, the maximum temperature evolution displays an exponential decay predicted by the decay rate of edge mode [$-{\rm{Im}}(\omega)_{\rm{edge}}=0.101$~rad/s, see Fig.~\ref{Fig4}(d)]. However, in the trivial phase, the temperature evolution diverges from the nontrivial result, failing to exhibit an exponential decay [see Fig.~\ref{Fig4}(d)]. Additionally, the maximum temperature evolution of the edge ring remains robust against disorder of the rotating velocities of rings [see Appendix~\ref{app:AppC} and Fig.~\ref{Fig8}(b)]. The temperature field without convection modulation shows an similar distribution with the one in the trivial phase [see Appendix~\ref{app:AppE} and Fig.~\ref{Fig10}(c)]. These findings further demonstrate the unique thermal properties of topological edge modes in our system.

\section{\label{sec:extend_local}Convection-modulated extended-localized criticality} 

\begin{figure}
\includegraphics[width=0.9\linewidth]{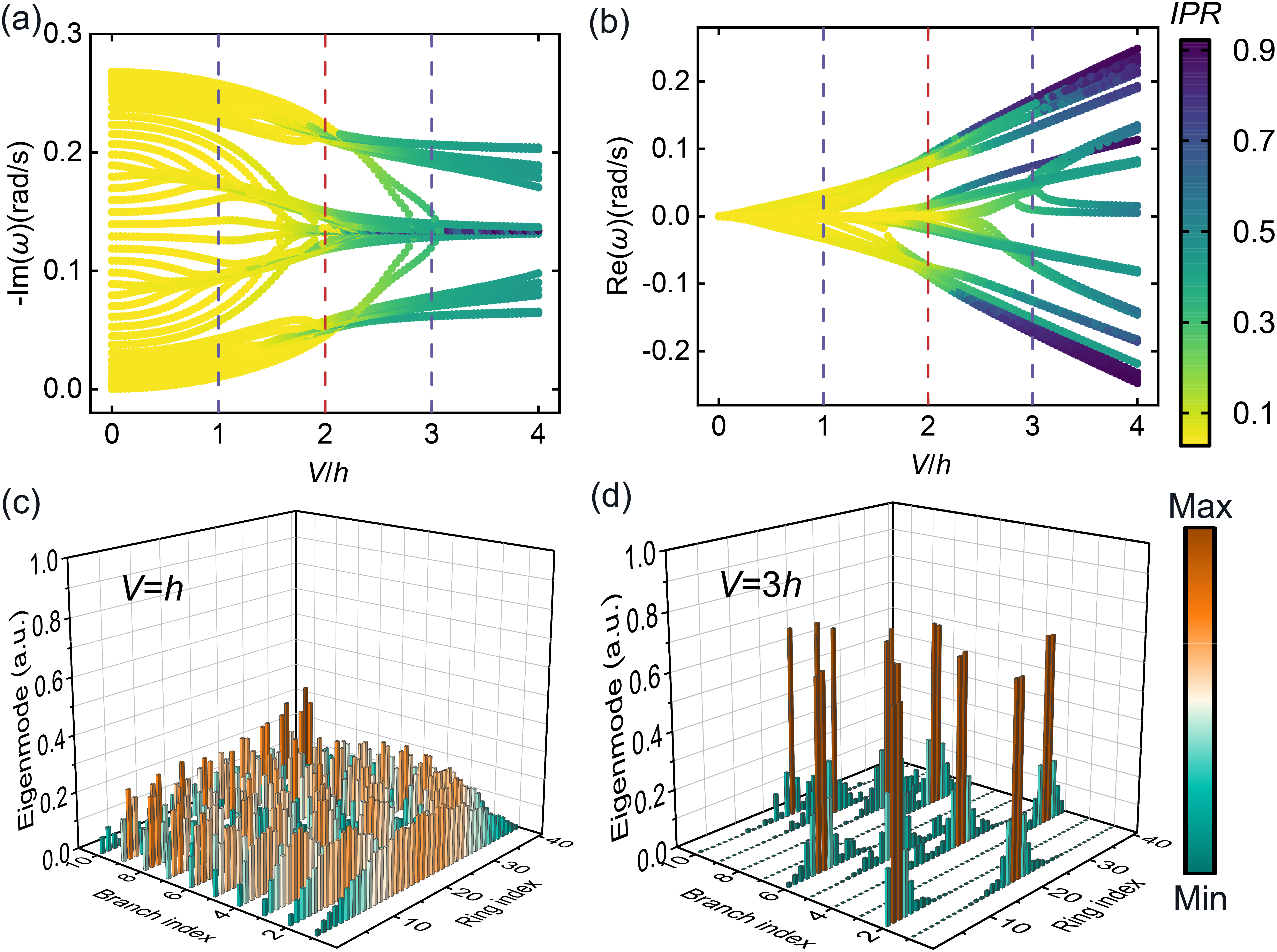}
\caption{Eigenvalue and eigenstate of diffusive model when $\alpha=(\sqrt{5}-1)/2$. (a,b) The theoretical (a) imaginary and (b) real spectra with different amplitudes of imaginary onsite potential $V$. The colorbar indicates the $IPR$. The dashed red line marks the extended-localized transition point $V=2h$. (c,d) Eigenstate distributions of the ten slowest decaying branches for (c) the extended state ($V=h$) and (d) the localized state ($V=3h$). Both of these two states are marked as dashed purple lines in Figs.~\ref{Fig5}(a) and (b). Here $\delta=0.4\pi$.}
\label{Fig5}
\end{figure}

In this section, we will study the extended-localized transition induced by convection modulation. Due to its self-duality, the quasiperiodic Hermitian AAH model is known to have an extended-localized transition at a finite strength of onsite potential. Here, we incorporate the quasiperiodicity into our diffusive theoretical model with $\alpha=(\sqrt{5}-1)/2$. Figures~\ref{Fig5}(a) and (b) clearly demonstrate an extended-localized transition at $V=2h$ in both the imaginary and real spectra, indicating that non-Hermitian quasiperiodic model inherits the localization nature of Hermitian AAH model. To quantify the degree of localization for each eigenstate, we utilize the inverse participation ratio ($IPR$)~\cite{LonghiPRL19, LonghiPRB19, JiangPRB19, LiuPRB21-1, LiuPRB21-2}. The $IPR$ is defined as $IPR=\sum_{j}|\psi_{j}(E)|^4/(\sum_{j}|\psi_{j}(E)|^2)^2$, where $\psi_{j}(E)$ is the $j$-th component of the eigenstate corresponding to energy $E$. For an extended state, the $IPR$ approaches zero as $N{\rightarrow}\infty$ ($IPR{\approx}1/N$), whereas for a fully localized state, the $IPR$ is approximately one. The eigenstate distributions of the ten slowest decaying branches for the extended state ($V=h$) and the localized state ($V=3h$) are shown in Figs.~\ref{Fig5}(c) and (d), respectively. The extended state exhibits a nearly uniform eigenstate distribution, while the localized state shows localization at specific rings. Additionally, we have extracted the Lyapunov exponent of non-Hermitian quasiperiodic model from the eigenstate distributions, as discussed in Appendix~\ref{app:AppD}. A clear extended-localized transition is also observed in the Lyapunov exponent (see Fig.~\ref{Fig9}).

\begin{figure}
\includegraphics[width=\linewidth]{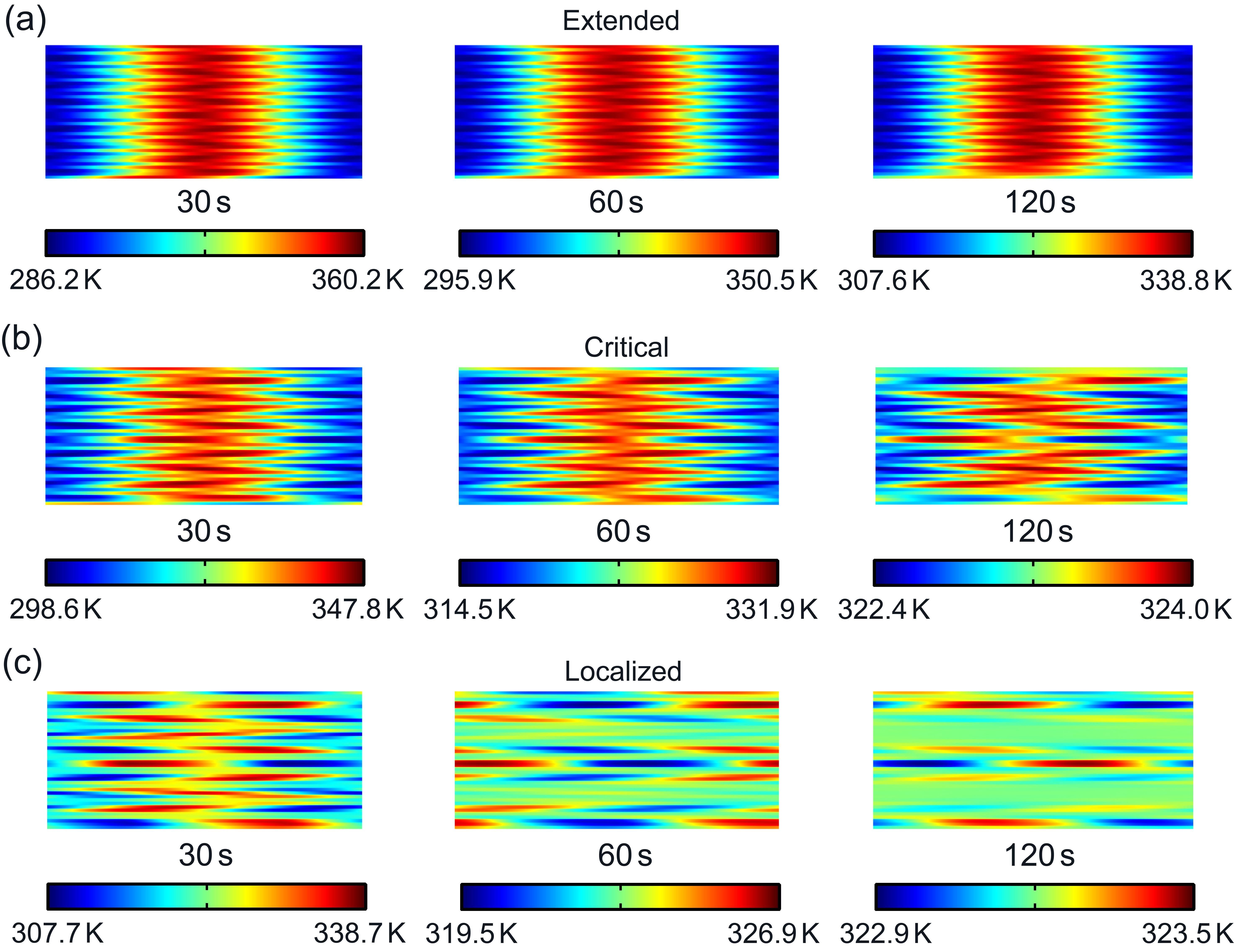}
\caption{Temperature field simulations with $\alpha=(\sqrt{5}-1)/2$. (a) Extended state ($V=h$), (b) critical state ($V=2h$), and (c) localized state ($V=3h$).}
\label{Fig6}
\end{figure}

Next, we perform the temperature field simulation to explore the extended-localized transition (see the method in Appendix~\ref{app:AppB}). For the diffusive extended state at $V=h$, the temperature field remains evenly distributed over time, as shown in Fig.~\ref{Fig6}(a). The temperature profile remains stationary due to the relatively small real part of the eigenvalues [see Fig.~\ref{Fig5}(b)]. In contrast, for the diffusive localized state at $V=3h$ [see Fig.~\ref{Fig6}(c)], the temperature distribution becomes highly localized with multiple distinct centers. This multiplicity arises from the closeness of decay rates among several slow decaying branches and is unique to diffusion systems without wave counterpart~\cite{LiuPRAP24, LiuCPL23}. Additionally, the temperature field moves drastically due to the larger real spectrum in the localized phase [see Fig.~\ref{Fig5}(b)]. The moving double localization centers (moving multiple localization centers can be generalized naturally) phenomenon here would help design a potential application dubbed as double-trace generator [see Appendix~\ref{app:AppF} and Fig.~\ref{Fig11}]. For the critical state at $V=2h$, the temperature field lies between the extended and localized states [see Fig.~\ref{Fig6}(b)]. Without any convection modulation, the thermal field shows a perfectly uniform distribution [see Appendix~\ref{app:AppE} and Fig.~\ref{Fig10}(e)].  

\section{\label{sec:conclusion}Conclusion}

We have shown that convections in thermal metamaterials can be harnessed to engineer topological edge modes and localization transition. Utilizing the coupled ring chain structure, we have realized the tight-binding model with a modulated imaginary onsite potential in condensed matter physics. Under a periodic convection modulation, robust edge modes emerge within the band gap with an exponentially localized distribution. The temperature field of these edge states is localized and decays exponentially. Besides, we observe an extended-localized criticality in the quasiperiodic limit. In the localized phase, the temperature field displays a motioning multiple localization centers phenomenon. From the view of application, this convective modulation has a very flexible adjustability. Thermal topological edge mode has a potential application in the localized heat management and robust heat transport. Our findings offer a versatile platform for engineering topological modes and criticality through convection modulation, inspiring further researches in reconfigurable thermotics.

\appendix

\section{\label{app:AppA}Non-Hermitian electric polarization as the topological invariant}

\begin{figure}
\includegraphics[width=0.5\linewidth]{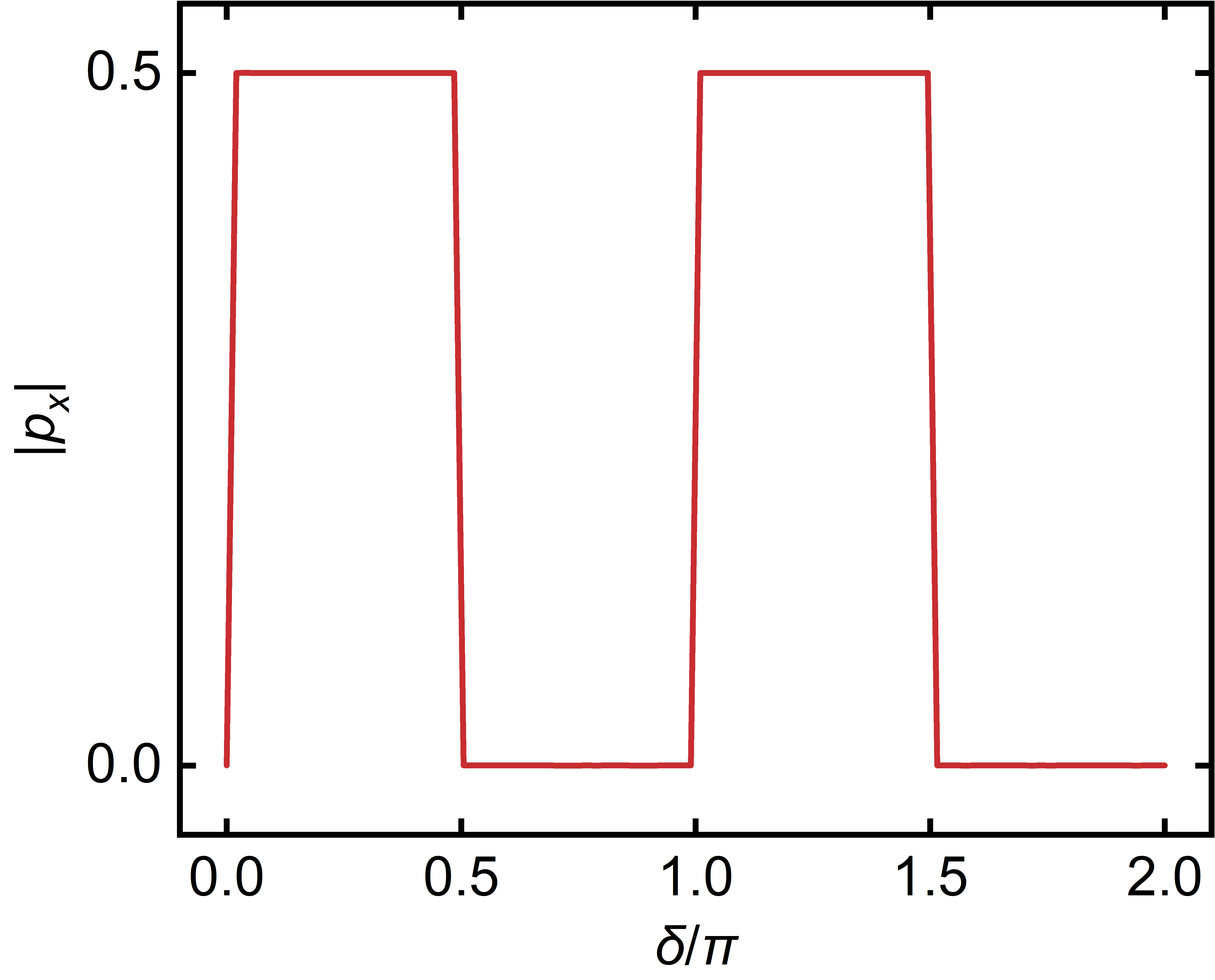}
\caption{Non-Hermitian electric polarization $|p_{x}|$ with different phases $\delta$.}
\label{Fig7}
\end{figure}

In this section, we calculate the non-Hermitian electric polarization to characterize the topological properties of diffusive theoretical model~\cite{ZhuPRR23}. Firstly, as the Hamiltonian is non-Hermitian, the eigenvectors obey a biorthogonality relation ${\langle}u_{m}^{L}(k)|u_{n}^{R}(k){\rangle}={\delta}_{mn}$, where $m$, $n$ are occupied band indices, $u^{L}$ ($u^{R}$) is the left (right) eigenvector of Bloch Hamiltonian, and $\delta_{mn}$ is the Kronecker symbol ($\delta_{mn}=1$ when $m=n$ and $\delta_{mn}=0$ when $m{\neq}n$). Then we adopt the Wilson loop method to calculate the polarization. The non-Abelian Wilson loop operator is a $(N/2){\times}(N/2)$ matrix, which is defined as:
\begin{equation}\label{5}
W=P_{k+2{\pi}-{\rm{\Delta}}k}{\cdot}P_{k+2{\pi}-2{\rm{\Delta}}k}{\cdots}P_{k+{\rm{\Delta}}k}{\cdot}P_{k},
\end{equation} 
where 
\begin{equation}\label{6}
\left[P_{k}\right]^{mn}={\langle}u_{m}^{L}(k)|u_{n}^{R}(k+{\rm{\Delta}}k){\rangle}.
\end{equation} 
The electric polarization is related to the Wilson loop by
\begin{equation}\label{7}
p_{x}=\frac{i}{2{\pi}}{\rm{ln}}({\rm{det}}{\,}[W]).
\end{equation} 
The result of this polarization is shown in Fig.~\ref{Fig7}. In the topological nontrivial phase, the polarization is quantized with $|p_{x}|=1/2$; while in the trivial phase, $p_{x}=0$. These results obtained from the periodic boundary condition can predict the presence or absence of topological edge modes under the open boundary condition.

\section{\label{app:AppB}Simulation method}

In this work, we use the commercial software COMSOL Multiphysics to perform the simulation. We adopt the heat transfer module. The periodic boundary condition is imposed at the ends of each channel. For the simulated eigenvalues and eigenstates shown in Fig.~\ref{Fig3}, we perform the eigenvalue simulation on the coupled ring chain structure. For the temperature field simulation shown in Figs.~\ref{Fig4} and~\ref{Fig6}, we perform the transient state simulation on the whole structure. The initial condition used in Fig.~\ref{Fig6} is the uniform excitation, which means that the initial temperature linearly distributes between the middle (highest temperature $T_h$) and the end (lowest temperature $T_l$) of the channel. Parameters used in this work are shown as follows. The thickness of ring is $b=2$~mm. The thickness of interlayer is $d=0.5$~mm. The radius of ring is $R=50$~mm. The thermal conductivity of ring is $\kappa=1.2$~W/(m$\cdot$K). The mass density of ring is $\rho=1000$~kg/m$^3$. The heat capacity of ring is $C=1640$~J/(kg$\cdot$K). The thermal conductivity of interlayer is $\kappa_{\rm{I}}=0.11$~W/(m$\cdot$K). The number of rings is $N=40$. The highest and lowest temperatures in the simulation are set as $T_{h}=373.2$~K and $T_{l}=273.2$~K, respectively. In the $y$-axis label in Figs.~\ref{Fig4}(d) and~\ref{Fig8}(b), $T_{f}=(T_{h}+T_{l})/2=323.2$~K and $T_{i}=T_{h}=373.2$~K.

\section{\label{app:AppC}Robustness against disorder on topological edge mode}

\begin{figure}
\includegraphics[width=0.9\linewidth]{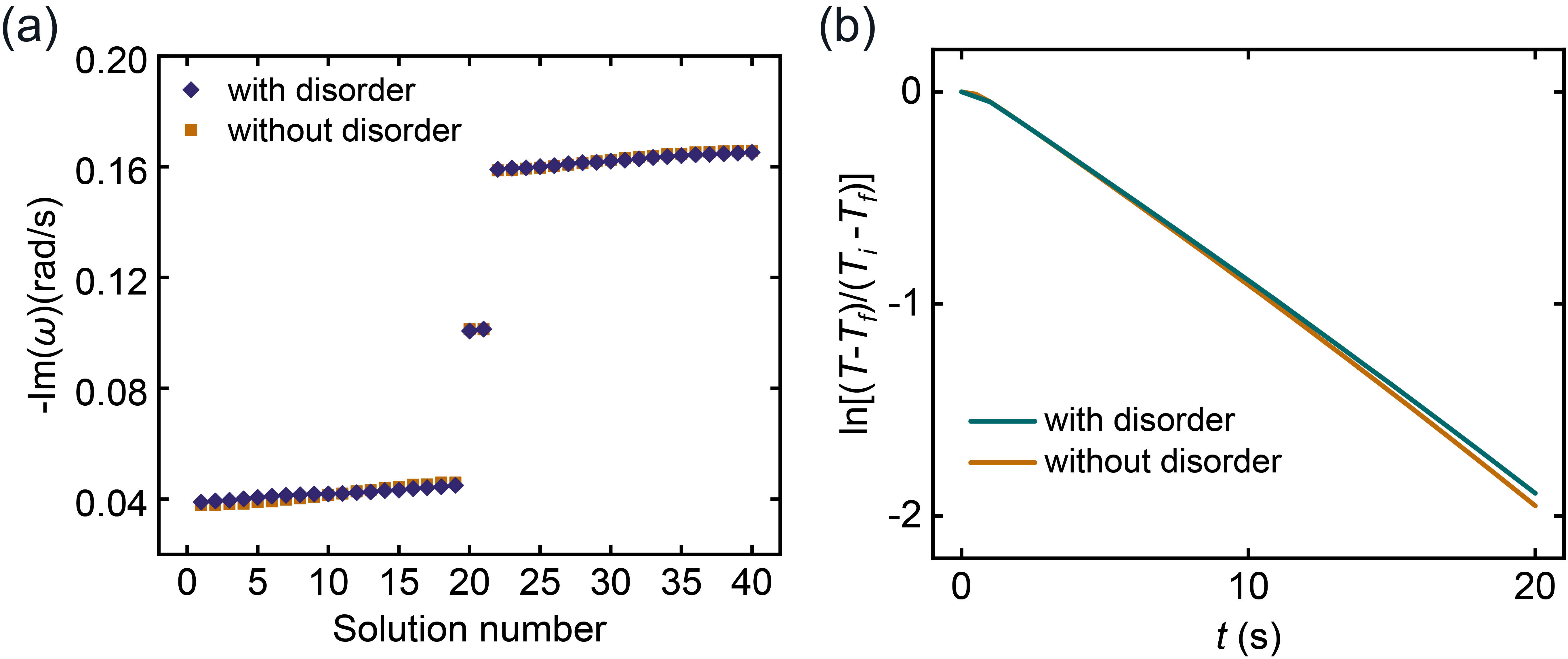}
\caption{The influence of disorder on the edge state. (a) The simulated imaginary spectrum at $\delta=0.25\pi$ with and without velocity disorder. (b) The normalized maximum temperature evolution ${\rm{ln}}[(T-T_{f})/(T_{i}-T_{f})]$ of the edge (1st or 40th) ring at $\delta=0.25\pi$ with and without velocity disorder. Here $V_{D}=h$ and the times of sampling is 50.} 
\label{Fig8}
\end{figure}

In this section, we study the influence of disorder on topological edge modes. The disorder is included in the theoretical model Hamiltonian by adding a term 
\begin{equation}\label{8}
H_{D}=iV_{D}\sum\limits_{j=1}^{N}{\chi}_{j,D}a_{j}^{\dagger}a_{j},
\end{equation} 
where $V_{D}$ is the disorder strength and ${\chi}_{j,D}$ is a Gaussian distribution function with average 0 and width 1. The effect of this disorder will reflect on the rotating velocity of each ring. After a certain number of samples, we can study the effect of disorder on the thermal structure. In the simulated imaginary spectrum [see Fig.~\ref{Fig8}(a)], we can find that two topological edge modes remain localized in the band gap with an unaltered decay rate despite of the disorder. Besides, choosing the edge mode distribution as the initial condition, the temperature evolution of edge ring keeps almost unchanged in the presence of disorder [see Fig.~\ref{Fig8}(b)]. Therefore, the topological edge mode here is robust against the disorder of rotating velocity.

\section{\label{app:AppD}Lyapunov exponent}

\begin{figure}
\includegraphics[width=0.5\linewidth]{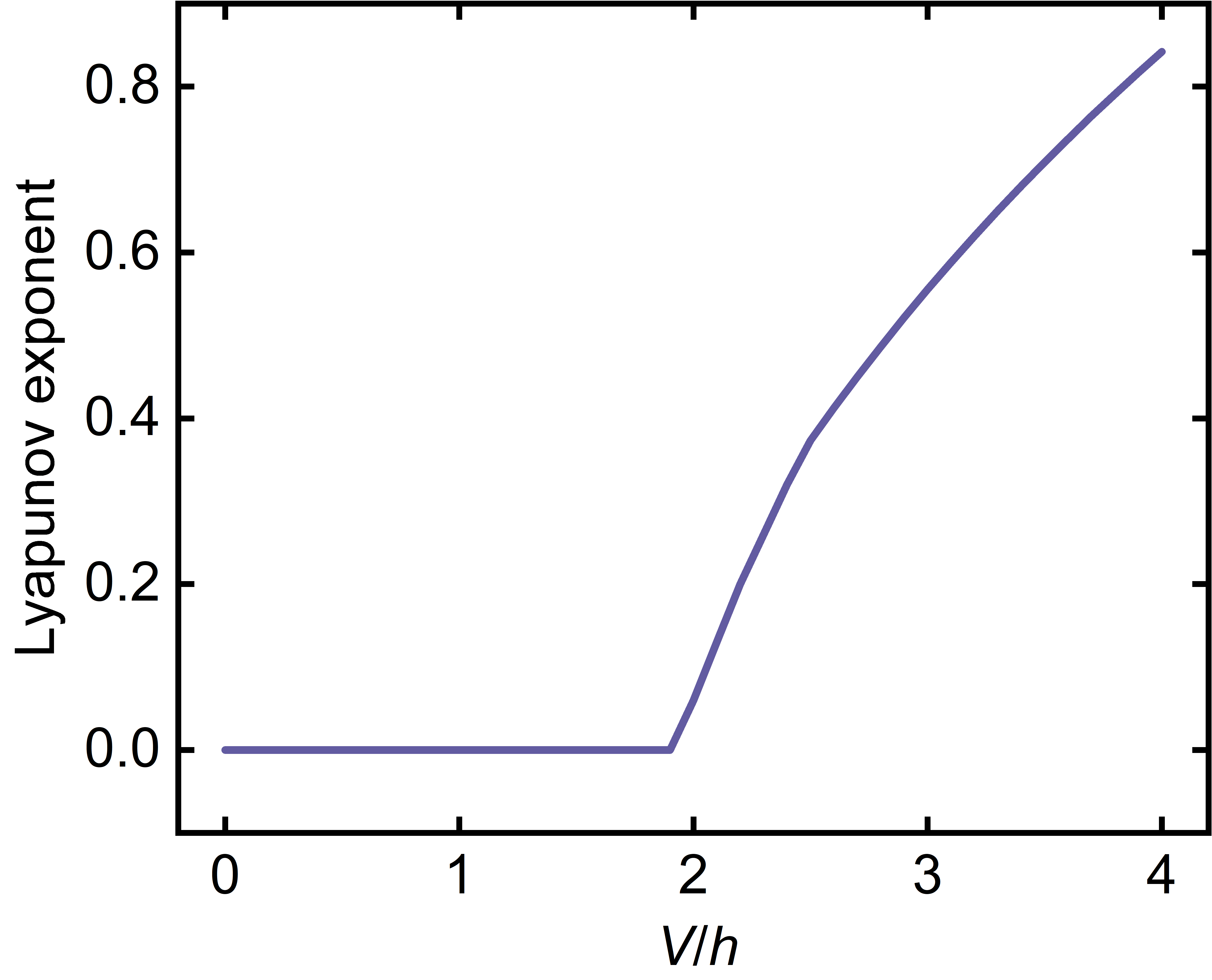}
\caption{Theoretical Lyapunov exponent with different amplitudes of onsite potential $V$.}
\label{Fig9}
\end{figure}

Here we extract the Lyapunov exponent of theoretical non-Hermitian AAH model, which can also be used to study the extended-localized transition. In the localized phase, the eigenstate has an exponential form with $|\psi|=|\psi|_{\rm{max}}e^{-{\lambda}|i-i_{0}|}$, where $i_{0}$ is the ring index of the localization center and $\lambda$ is the Lyapunov exponent~\cite{JiangPRB19}. We perform an exponential fit to the eigenstate distribution and then extract the Lyapunov exponent (see Fig.~\ref{Fig9}). The Lyapunov exponent for the extended state is zero. For the localized state, the Lyapunov exponent increases with the amplitude of onsite potential, so the state becomes more localized. From the Lyapunov exponent, we can find that a clear extended-localized transition occurs at $V=2h$.

\section{\label{app:AppE}The case without convection modulation}

\begin{figure}
\includegraphics[width=\linewidth]{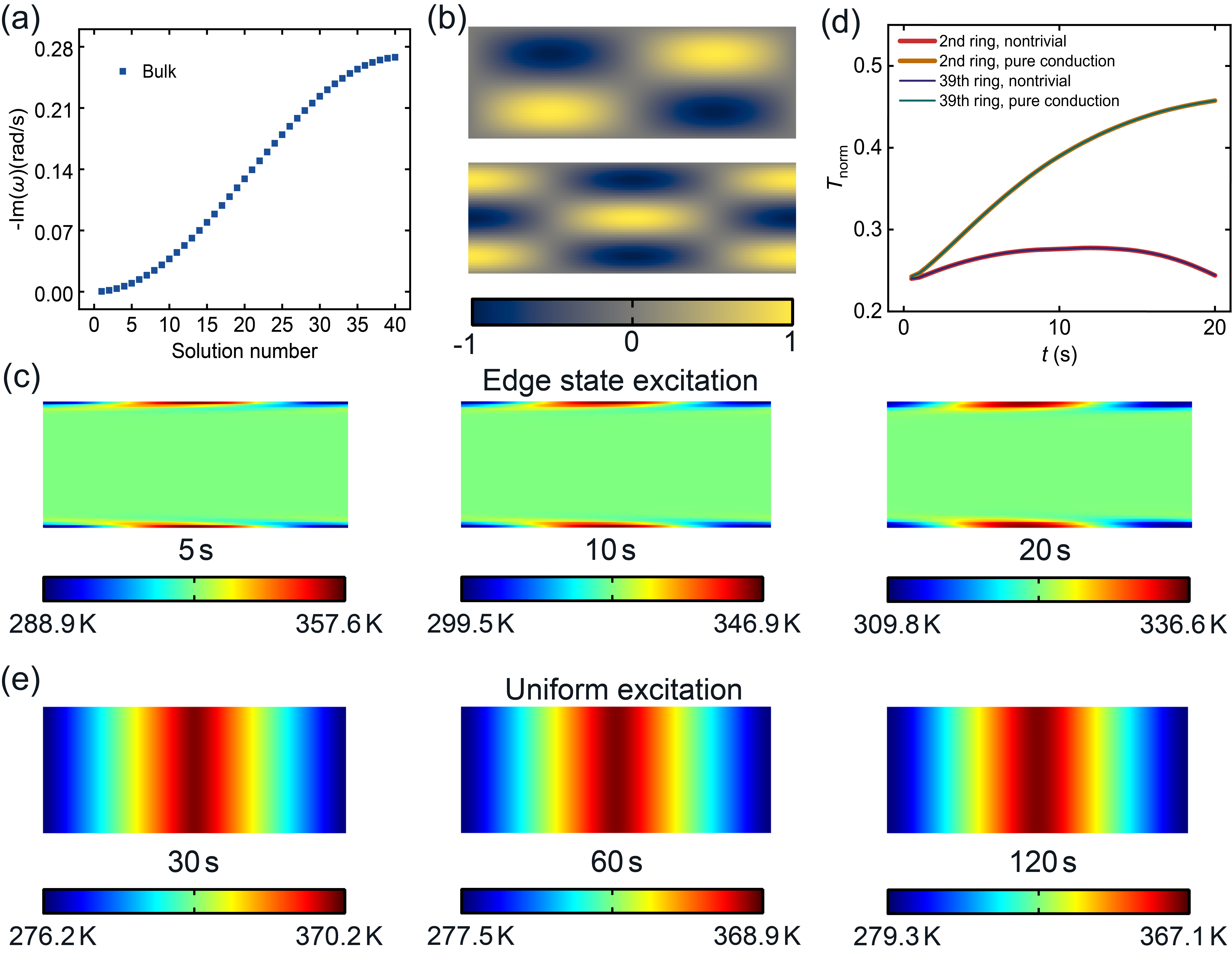}
\caption{The case without convection modulation. (a) The theoretical imaginary spectrum. The blue square indicates the bulk mode. (b) The simulated profiles of two eigenmodes. (c) Temperature field simulation with edge state excitation. (d) The normalized temperature field $T_{\rm{norm}}$ for the 2nd and 39th rings with time evolution without convection modulation. We also plot the result in the nontrivial phase for comparison, which is extracted in the Fig.~\ref{Fig4}(c). (e) Temperature field simulation with uniform excitation.}
\label{Fig10}
\end{figure}

In this appendix, we study the case without convection modulation. Firstly, we can find that there is no band gap in the imaginary spectrum [see Fig.~\ref{Fig10}(a)]. Besides, the distributions for two representative eigenstates are uniform without any localization [see Fig.~\ref{Fig10}(b)]. Then we perform the temperature field simulations without any convection modulation. Corresponding to the main text, we choose two initial conditions to study this issue: one is setting the distribution of topological edge state ($\delta=0.25\pi$) as the initial excitation and the other is the uniform excitation. For the edge state excitation, the temperature field without convection modulation [see Fig.~\ref{Fig10}(c)] has a similar distribution with the one in the trivial phase [$\delta=0.75\pi$, see Fig.~\ref{Fig4}(b)]. It shows a weaker localization at the edge rings than the one in the nontrivial phase [$\delta=0.25\pi$, see Fig.~\ref{Fig4}(a)]. For a quantitative analysis, we extract the normalized temperature field for the 2nd and 39th rings without convection modulation [see Fig.~\ref{Fig10}(d)]. Compared with the one in the nontrivial phase, the normalized temperature field without convection modulation increases with time evolution. For the uniform excitation, the temperature field without convection modulation shows a perfectly uniform distribution [see Fig.~\ref{Fig10}(e)], which exhibits a totally different behaviour with the case for $V=h$, $V=2h$, and $V=3h$ (see Fig.~\ref{Fig6}). Therefore, it is plausible to conclude that all the exotic phenomena observed in the temperature field within this study can be attributed to the convection modulation.

\section{\label{app:AppF}Potential application: double-trace generator}

\begin{figure}
\includegraphics[width=0.6\linewidth]{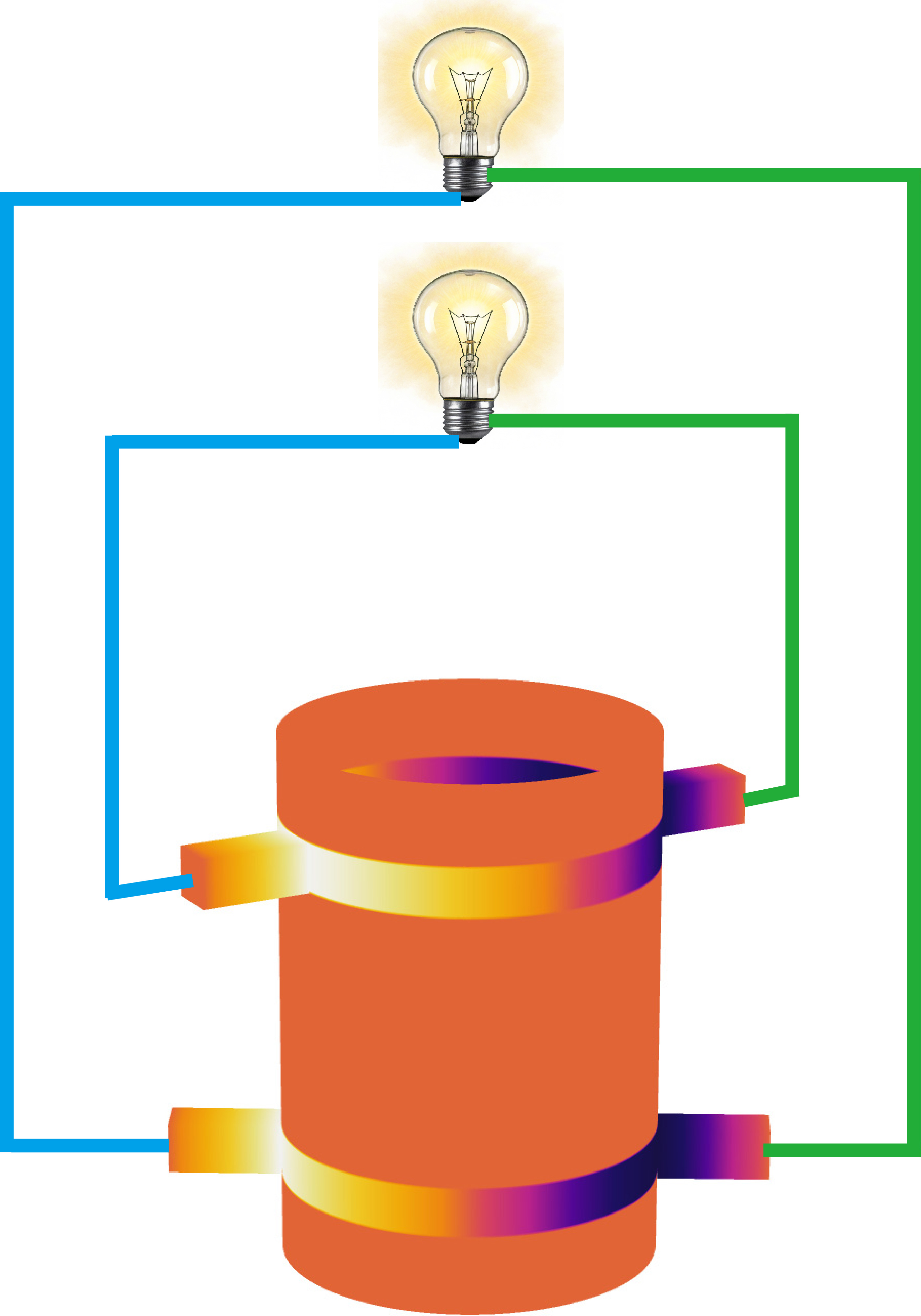}
\caption{Double-trace generator. The cuboid indicates the thermoelectric material. The temperature at the hot and cold sources are coloured in white and modena, respectively.}
\label{Fig11}
\end{figure}

In this appendix, we propose a potential application dubbed as double-trace generator based on the moving double localization centers (moving multiple localization centers can be generalized naturally). Compared with the case without convection modulation, it exhibits an extended-localized transition after introducing the convection with a quasiperiodic modulation. Especially, a moving multiple localization centers phenomenon will emerge in the localized phase. This moving multiple localization centers phenomenon can be clearly observed in the temperature field. Here we consider the case of moving double localization centers for simplicity (see Fig.~\ref{Fig11}). After contacting with the double localization centers, the two pairs of thermoelectric materials would convert thermal energy into electricity for double-trace generation (such as lighting two light bulbs simultaneously). Here the dynamical double localization centers are achieved by adjusting the rotating velocity of each ring. In condensed matter physics, most of the non-Hermitian quasicrystals studied have a real or complex onsite potential~\cite{LonghiPRL19, LonghiPRB19, JiangPRB19, LiuPRB21-1, LiuPRB21-2, WangPRL20, ZhouPRL23}. The realization of these models in thermotics needs to adjust the thermal conductivity of each ring. The tunability of thermal conductivity is more difficult and less flexible than the one of rotating velocity. So the double-trace generator proposed in this work can be implemented easier, which is a reconfigurable thermal device.

\section*{Acknowledgements}
Y. L. acknowledges the support by the National Key Research and Development Program of China under Grant No. 2023YFB4604100, the National Natural Science Foundation of China under Grants Nos. 12475040, 92163123 and 52250191, and Zhejiang Provincial Natural Science Foundation of China under Grant No. LZ24A050002. J. H. acknowledges the financial support provided by the National Natural Science Foundation of China under Grants No. 12035004 and No. 12320101004, and the Innovation Program of the Shanghai Municipal Education Commission under Grant No. 2023ZKZD06.

\end{document}